\documentclass[aps,superscriptaddress,eqsecnum,nofootinbib,showpacs,preprintnumbers]{revtex4}
\usepackage{graphicx}
\usepackage{amsmath}
\usepackage {amssymb}

\begin{document}

\title{Induced gravity with a non-minimally coupled scalar field on the brane}

\author{Mariam Bouhmadi-L\'opez}\email{mariam.bouhmadi@port.ac.uk}
\affiliation{Institute of Cosmology and Gravitation, University of
Portsmouth,  Mercantile House, Hampshire Terrace  Portsmouth  PO1
2EG, UK}

\author{David Wands}\email{david.wands@port.ac.uk}
\affiliation{Institute of Cosmology and Gravitation, University of
Portsmouth,  Mercantile House, Hampshire Terrace  Portsmouth  PO1
2EG, UK}

\date{August 6, 2004}

\begin{abstract}

  We present the four-dimensional equations on a brane with a scalar
  field non-minimally coupled to the induced Ricci curvature, embedded
  in a five-dimensional bulk with a cosmological constant. This is a
  natural extension to a brane-world context of scalar-tensor
  (Brans-Dicke) gravity. In particular we consider the cosmological
  evolution of a homogeneous and isotropic (FRW) brane.  We identify
  low-energy and strong-coupling limits in which we recover
  effectively four-dimensional evolution. We find de Sitter brane
  solutions with both constant and evolving scalar field. We also
  consider the special case of a conformally coupled scalar field for
  which it is possible (when the conformal energy density exactly
  cancels the effect of the bulk black hole) to recover a conventional
  four-dimensional Friedmann equation for all energy densities.

\end{abstract}

\pacs{04.50.+h, 98.80.-k  \ \hfill hep-th/0408061}

\maketitle

\section{Introduction}

Over recent years there has been a great deal of interest in
higher-dimensional models of space-time where matter fields are
restricted to a lower-dimensional brane in a higher-dimensional bulk
space-time: the simplest case being a 3-brane of codimension one in a
five-dimensional (5D) bulk. 

This raises the possibility that the
four-dimensional (4D) gravity we observe is the projection of a
higher-dimensional gravity. In particular Randall and Sundrum
\cite{RS2} discovered that conventional 4D gravity can be recovered at
large scales (low energies) on a Minkowski brane-world embedded in a
5D anti-de Sitter space-time.
Even if there is no 4D Einstein-Hilbert term in the classical
theory then such a term should be induced by loop-corrections from
matter fields \cite{Collins}. Dvali, Gabadadze and Porrati
\cite{DGP} argued that in this case 4D gravity can then be recovered
at small scales (high energies) on a Minkowski brane-world in 5D
Minkowski space-time.
More generally one can consider the effect of an induced gravity term
as a quantum correction in any brane-world model such as the
Randall-Sundrum model.

Cosmology is a natural arena in which to put to the test alternative
theories of gravity. In particular the DGP model admits late-time
accelerating solutions. The cosmology of induced gravity corrections
to Randall-Sundrum type models have been considered by several
authors~\cite{Kofinas,Deffayet,Kiritsis,MMT,lefteris,BLMW}.

In this paper we will consider the effect of an induced gravity term
which is an arbitrary function of a scalar field on the brane.
Scalar fields play an important role both in models of the early
universe and late-time acceleration. They also provide a simple
dynamical model for matter fields in a brane-world model. In the
context of induced gravity corrections it is then natural to
consider a non-minimal coupling of the scalar field to the intrinsic
(Ricci) curvature on the brane that is a function of the field.
The resulting theory can be thought of as a generalisation of
Brans-Dicke type scalar-tensor gravity in a brane-world context.

The layout of this paper is as follows.  In section II we present the
five- and four-dimensional terms in the action and then use the
geometrical approach of Shiromizu, Maeda and Sasaki \cite{SMS} to give
the effective Einstein equations projected onto the brane. Although in
general these equations are not closed, due to the presence of the
projected 5D Weyl tensor, the symmetries of a homogeneous and
isotropic brane cosmology are sufficient to determine the evolution of
the projected Weyl tensor on the brane \cite{Kraus,BDEL,BCG}. In
section III we identify two regimes in which we expect to recover
effectively 4D behaviour and in section IV we show that this is indeed
the case for cosmological (homogeneous and isotropic) branes. We
discuss static (de Sitter or Minkowski) brane solutions in section V
and then consider the special case of a conformally coupled scalar
field on the brane in section VI. The rather complicated form of the
modified Friedmann equation on the brane is somewhat simpler for a
conformally coupled field and we show that it is even possible to
recover a conventional four-dimensional Friedmann equation, at all
energies, as a special case. Finally we summarise our results in
section VII.

\section{Induced scalar-tensor gravity action}

\subsection{5D gravity}

We consider a brane, described by a 4D hypersurface ($b$ , metric
$g$), embedded in a  5D bulk space-time ($\mathcal{B}$, metric
$g^{(5)}$), whose action is given by
\begin{eqnarray}
S = \int_{\mathcal{B}} d^5X\, \sqrt{-g^{(5)}}\;
\left\{\frac{1}{2\kappa_5^2}R[g^{(5)}]+\mathcal{L}_5\right\} +
\int_{b} d^4X\, \sqrt{-g}\; \Big\{
\frac{1}{\kappa_5^2} K +
\mathcal{L}_4\Big\} \,,
 \label{action}
\end{eqnarray}
where $\kappa_5^2$ is the 5D gravitational constant, $R[g^{(5)}]$
is the Ricci scalar in the bulk and $K$ the extrinsic curvature of
the brane in the higher-dimensional bulk, corresponding to the
York-Gibbons-Hawking boundary term \cite{YGH}. Thus we have 5D
Einstein gravity with a 4D boundary.

We will consider the simplest case of a constant vacuum energy density
on the bulk, $\mathcal{L}_5=-U$, i.e., a cosmological constant.  In
this case the bulk geometry is given by an Einstein space with
constant scalar curvature
\begin{equation}
G_{MN}[g^{(5)}]=-\kappa_5^2 U g_{MN}^{(5)} \,.
\end{equation}

\subsection{4D induced gravity}

For simplicity we will assume a $\mathbb{Z}_2$-symmetry at the
brane (which is also motivated by specific M-theory constructions
\cite{HW,LOW}). In practice one can easily generalise to
non-$\mathbb{Z}_2$-symmetric branes \cite{carter}. The effective
Einstein equation on the brane is then \cite{SMS}
\begin{equation}
G_{\mu\nu}[g]=-\frac12 \kappa_5^2 U g_{\mu\nu} +\kappa_5^4
{{\Pi}}_{\mu\nu}-E_{\mu\nu}, \label{Einsteineq}\end{equation}
where $g$ is the induced metric on the brane. ${{\Pi}}_{\mu\nu}$
is the quadratic energy momentum tensor~\cite{SMS}
\begin{eqnarray}
{{\Pi}}_{\mu\nu}=-\frac14
\tau_{\mu\sigma}{\tau_{\nu}}^{\sigma}+\frac{1}{12}\tau
\tau_{\mu\nu}+\frac{1}{8}
g_{\mu\nu}(\tau_{\rho\sigma}\tau^{\rho\sigma}-\frac{1}{3}\tau^2)
\,, \label{quadratic}\end{eqnarray}
and $\tau_{\mu\nu}$ is the total energy-momentum tensor for fields on
the brane defined by
\begin{eqnarray}
\tau_{\mu\nu}&=&-2\frac{\delta {\mathcal{L}}_4}{\delta
g^{\mu\nu}}+ g_{\mu\nu}{\mathcal{L}}_4 \,.
 \label{deftau}
\end{eqnarray}
$E_{\mu\nu}$ is the (trace-free) projected Weyl tensor on the
brane. The trace-free property determines the isotropic effective
pressure of this projected Weyl tensor in terms of its effective
density, but the anisotropic effective pressure due to this
non-local term cannot in general be determined without some
additional information about the 5D gravitational field.

The most general scalar field Lagrangian $\mathcal{L}_4$ for a scalar
field, $\phi$, confined on the brane can be written as
\begin{eqnarray}
\mathcal{L}_4=-\frac12 g^{\mu\nu} \nabla_{\mu}\phi\nabla_{\nu}\phi
-V(\phi) +\alpha(\phi) R[g],\label{actionphi}
\end{eqnarray}
where $\nabla_\mu$ is the covariant derivative associated with the
induced metric on the brane $g$. Previous studies of scalar fields in
induced brane-world gravity \cite{lefteris} are restricted to the
case $\alpha=$~constant.  Here we include a coupling between the
scalar field $\phi$ and the induced gravity term on the brane, given
by $\alpha(\phi)$.  In this case, substituting (\ref{actionphi}) into
(\ref{deftau}), the total energy-momentum tensor on the brane becomes
\begin{eqnarray}
 \label{altdeftau}
\tau_{\mu\nu}=\nabla_\mu\phi\nabla_\nu\phi-\frac{1}{2}g_{\mu\nu}(\nabla\phi)^2
-g_{\mu\nu}V(\phi)-2\alpha\mathcal{G}_{\mu\nu}[g,\phi] \,.
\end{eqnarray}
This includes the ``Einstein-Brans-Dicke'' tensor
\begin{eqnarray}
\mathcal{G}_{\mu\nu}[g,\alpha] \equiv G_{\mu\nu}[g]+
\frac{1}{\alpha}(g_{\mu\nu}g^{\rho\sigma}-{g_\mu}^{\rho}{g_\nu}^{\sigma})
(\alpha'\nabla_\rho\nabla_\sigma\phi+\alpha''\nabla_\rho\phi\nabla_\sigma\phi),
\label{EinsteinBD}\end{eqnarray}
due to the non-minimal coupling, $\alpha(\phi)$, between the scalar field
$\phi$ and the scalar curvature $R[g]$. In this expression the
prime denotes derivative with respect to $\phi$.

We can thus split the total energy-momentum tensor as follows
\begin{equation}
 \label{deftausplit}
\tau_{\mu\nu}=
{T_{\mu\nu}}^{(\phi)}+{T_{\mu\nu}}^{(\alpha)}-2\alpha
G_{\mu\nu}[g],
\end{equation}
where the canonical (minimally coupled) scalar field energy-momentum
tensor is given by
\begin{eqnarray}
 \label{Tphi}
{T_{\mu\nu}}^{(\phi)} \equiv
 \nabla_\mu\phi\nabla_\nu\phi-\frac{1}{2}g_{\mu\nu}(\nabla\phi)^2
-g_{\mu\nu}V(\phi),
\end{eqnarray}
and the extra terms arising from the dependence of the induced gravity
term upon $\phi$ are given by
\begin{eqnarray}
{T_{\mu\nu}}^{(\alpha)}\equiv
 - 2\left(g_{\mu\nu}g^{\rho\sigma}-{g_\mu}^{\rho}{g_\nu}^{\sigma}\right)
\left(\alpha'\nabla_\rho\nabla_\sigma\phi+\alpha''\nabla_\rho\phi\nabla_\sigma\phi\right).
\label{Tphialpha}\end{eqnarray}

Using the 5D Codacci equation one can show that the total
energy-momentum tensor $\tau_{\mu\nu}$ must be conserved on the
brane \cite{SMS}
\begin{equation}
\nabla^\nu\tau_{\mu\nu}=0. \label{conservationtau}\end{equation}

\subsection{Scalar field wave equation}

Finally, the equation of motion for the scalar field reads
\begin{equation}
\nabla^\mu\nabla_\mu\phi=V'-\alpha'R[g] \,.
\label{motiongeneralphi}\end{equation}
This is the same as the standard equation of motion for a
non-minimally coupled scalar field in 4D, but it is often
re-written using the Einstein-Brans-Dicke equations to give $R$ in
terms of the trace of the energy-momentum tensor. Here we must
take the trace of the effective Einstein equations on the brane
(\ref{Einsteineq}) to give
\begin{equation}
R = 2\kappa_5^2 U -\kappa_5^4 \Pi_\mu^\mu \,,
\end{equation}
where
\begin{equation}
\Pi_\mu^\mu = \frac14 \tau_{\mu\nu}\tau^{\mu\nu} - \frac{1}{12}\tau^2 \,.
\end{equation}

Although the wave equation (\ref{motiongeneralphi}) is sufficient to
determine to evolution of the scalar field $\phi$ given the induced
metric on the brane, the effective Einstein equation
(\ref{Einsteineq}) is not in general sufficient to determine the
evolution of the induced metric given $\phi$. This is due to the
presence of the non-local term $E_{\mu\nu}$, representing the bulk
gravitational field. Nonetheless if we restrict our analysis to
homogeneous and isotropic brane-worlds these symmetries restrict the
bulk solution to either (anti-)de Sitter or Schwarzschild-(anti-)de
Sitter and the equations become closed \cite{BCG}.

\section{Equations in low energy and strong coupling limits}

\subsection{Low-energy limit}

In order to obtain the effective Einstein equations
(\ref{Einsteineq}) in a low-energy limit close to the
Randall-Sundrum solution~\cite{RS2} it is helpful to define a
``renormalised'' energy-momentum tensor on the brane
\begin{equation}
 \label{bartau}
\bar\tau_{\mu\nu} = \tau_{\mu\nu} + \sigma g_{\mu\nu} \,.
\end{equation}
where $\sigma$ is a constant brane tension.
The quadratic tensor $\Pi_{\mu\nu}$ defined in Eq.~(\ref{quadratic})
then becomes
\begin{equation}
\Pi_{\mu\nu} = -\frac{1}{12} \sigma^2 g_{\mu\nu} + \frac16 \sigma
\bar\tau_{\mu\nu} + \bar\Pi_{\mu\nu} \,,
 \label{barPi}
\end{equation}
where $\bar\Pi_{\mu\nu}$ is the quadratic energy-momentum
tensor~(\ref{quadratic}) formed from $\bar\tau_{\mu\nu}$ instead of
$\tau_{\mu\nu}$.

Substituting Eq.~(\ref{barPi}) for $\Pi_{\mu\nu}$ into
Eq.~(\ref{Einsteineq}) gives
\begin{equation}
G_{\mu\nu}[g] = -\Lambda_4 g_{\mu\nu}
 + \frac{\kappa_5^4\sigma}{6} \bar\tau_{\mu\nu}
 +\kappa_5^4 \bar{{\Pi}}_{\mu\nu}-E_{\mu\nu} \,.
\label{ren-Einsteineq}
\end{equation}
where we have defined
\begin{equation}
\Lambda_4 = \frac{\kappa_5^2}{2} U + \frac{\kappa_5^4\sigma^2}{12}
\label{Lambda4}
 \,.
\end{equation}
For $U<0$ we can choose $\sigma=\sqrt{-6U/\kappa_5^2}$ so that
$\Lambda_4=0$, but in principle we can work with any value of $\sigma$ and
hence $\Lambda_4$.

The energy-momentum tensor on the right-hand-side of
Eq.~(\ref{ren-Einsteineq}) includes a
contribution from the Einstein tensor, so ultimately we can re-write
the induced gravity equations on the brane as
\begin{equation}
 \label{Einsteinlo}
2\Phi_{\rm lo}\mathcal{G}_{\mu\nu}[g,\Phi_{\rm lo}]
 = - \frac{6\Lambda_4}{\kappa_5^4\sigma}  g_{\mu\nu}
 + \bar{T}^{(\phi)}_{\mu\nu}
 +\frac{6}{\kappa_5^4\sigma} \left(
 \kappa_5^4 \bar{{\Pi}}_{\mu\nu}-E_{\mu\nu} \right)
 \,,
\end{equation}
where $\mathcal{G}_{\mu\nu}[g,\Phi_{\rm lo}]$ is the
Einstein-Brans-Dicke tensor (\ref{EinsteinBD}) for the effective
Brans-Dicke field
\begin{equation}
 \label{defPhilo}
\Phi_{\rm lo}(\phi)
 \equiv \frac{3}{\kappa_5^4\sigma} \left[ 1 +
  \frac{\kappa_5^4\sigma}{3}\alpha(\phi)  \right] \,.
\end{equation}
Thus at low energies, if we can neglect the quadratic tensor
$\bar{{\Pi}}$, and in a conformally flat bulk ($E_{\mu\nu}=0$), we
will recover the usual Brans-Dicke equations for a non-minimally
coupled scalar field in four-dimensions. Moreover, for
$\alpha=$constant we recover Einstein gravity with a minimally
coupled scalar field on the brane and an effective gravitational
coupling $\kappa_4^2=(2\Phi_{\rm lo})^{-1}=$constant.

The effective potential for $\phi$ can be written as
\begin{equation}
V_{\rm lo}(\phi)
 = V(\phi) - \frac{\sigma}{2} + \frac{3U}{\kappa_5^2\sigma}
 \,,
\end{equation}
and the effective Brans-Dicke parameter is
\begin{equation}
\omega_{\rm lo} \equiv \frac{\Phi_{\rm lo}}{2[\Phi_{\rm lo}'(\phi)]^2}
 = \frac{3}{2\kappa_5^4\sigma\alpha^{\prime2}} \left[ 1 +
  \frac{\kappa_5^4\sigma}{3}\alpha(\phi)  \right] \,.
\label{omegalow}
\end{equation}

\subsection{Strong-coupling limit}

There is an alternative limiting case to consider where the 5D
curvature is negligible, or the induced coupling $\alpha$ is
large. In this case we expect the conventional 4D Lagrangian ${\cal
L}_4$ given in Eq.~(\ref{actionphi}) to dominate in the action
(\ref{action}). In this case we have the standard 4D
Einstein-Brans-Dicke equation
\begin{equation}
 \label{Einsteinhi}
2\alpha\mathcal{G}_{\mu\nu}[g,\alpha] = T^{(\phi)}_{\mu\nu} \,,
\end{equation}
with effective Brans-Dicke field
\begin{equation}
 \label{defPhihi}
\Phi_{\rm hi}(\phi) \equiv \alpha(\phi) \,,
\end{equation}
effective potential
\begin{equation}
V_{\rm hi}(\phi) = V(\phi) \,,
\end{equation}
and dimensionless Brans-Dicke parameter
\begin{equation}
\omega_{\rm hi}(\phi) = \frac{\alpha}{2\alpha^{\prime2}} \,.
\end{equation}
Note that this coincides with the limiting form of Eq.~(\ref{omegalow})
in the strong coupling limit, i.e., for $\kappa_5^4\sigma\alpha\gg1$.

\section{Dynamics of a homogeneous and isotropic brane}
\label{III}

In the present section we will consider the cosmological evolution
of a Friedmann-Robertson-Walker (FRW) brane with a non-minimally
coupled scalar field. In the special case of an isotropic brane
geometry the projected Weyl tensor $E_{\mu\nu}$ necessarily has a
vanishing anisotropic stress and the projected field equations
(\ref{Einsteineq}) and (\ref{motiongeneralphi}) form a closed set
of evolution equations for scalar field and metric on the brane.
Indeed, it can be shown that for an expanding FRW brane the unique
bulk space-time (in Einstein gravity in vacuum, as we assume here)
is 5D Schwarzschild-anti de Sitter space-time \cite{MSM,BCG}.

The trace-free property of the projected Weyl tensor implies that it
acts like a ``dark radiation'' \cite{Kraus,BDEL} and hence
\begin{equation}
{\dot E}_0^0+4HE_0^0=0 \,,
\label{B+C}
\end{equation}
where a dot denotes derivatives with respect to proper cosmic time and
$H$ is the Hubble rate. Thus $E_{\mu\nu}$ evolves like a radiation
fluid with $E_0^0=C/a^4$, where $C$ is an integration constant.

After some lengthy but straightforward calculations,
the modified Friedmann equation on the brane can be obtained from
Eq.~(\ref{Einsteineq}) as
\begin{eqnarray}
3\left(H^2+\frac{K}{a^2}\right)
 = \frac{\kappa_5^2U}{2}
 + \frac{\kappa^4_5}{12}\left[\rho-6\alpha
\left(H^2+\frac{K}{a^2}\right)\right]^2+\frac{C}{a^4}\;,
\label{Friedmann}
\end{eqnarray}
where $K=\pm 1,0$ depending on the geometry of the spatial
three-dimensional sections on the brane. The modified Friedmann
equation can be rewritten as
\begin{eqnarray}
 \label{Hubble}
H^2+\frac{K}{a^2}=\frac{1}{6\alpha}\left\{ \rho +
\frac{3}{\kappa_5^4\alpha}\left[1\pm\sqrt{1+\frac23\kappa_5^4\alpha\left(\rho-
\kappa_5^2\alpha U - 2\alpha\frac{C}{a^4}\right)}\right] \right\} \,,
\end{eqnarray}
which shows the existence of two branches of solution for $H^2$ as a
function of $\rho$.
The modified Raychaudhuri equation is
\begin{eqnarray}
\left\{1+ \frac{\kappa_5^4}{3}\alpha\left[\rho-6\alpha
\left(H^2+\frac{K}{a^2}\right)\right]\right\}\left(\dot
H-\frac{K}{a^2}\right)=-\frac{\kappa^4_5}{12}(\rho+P)\left[\rho-6\alpha
\left(H^2+\frac{K}{a^2}\right)\right]-\frac23\frac{C}{a^4}\;.
\label{Raychaudhuri}
\end{eqnarray}
Thus the modified Einstein equations can be written in exactly the
same form as obtained for constant $\alpha$ \cite{MMT}. The effect
of the non-minimal coupling of the $\phi$ field is hidden in the
definition of the effective energy density, $\rho$, of the scalar
field  which includes non-minimal terms. In the limit $\alpha\to0$
we recover the modified Einstein equations of the Randall-Sundrum
model \cite{BDEL} with a minimally coupled scalar field on the
negative branch (lower sign in Eq.(\ref{Hubble})).

Following the notation introduced in Eq.~(\ref{deftausplit}) we will write
\begin{eqnarray}
\rho &=& \rho^{(\phi)}+\rho^{(\alpha)} \,, \label{redrho}\\
P &=& P^{(\phi)}+ P^{(\alpha)} \,. \label{redp}
\end{eqnarray}
The effective energy density and pressure of the scalar field has
been split into a part associated with the canonical scalar field
energy-momentum tensor, given from Eq.~(\ref{Tphi}) as
\begin{eqnarray}
\rho^{(\phi)} &\equiv&
-{T^0_0}^{(\phi)} = \frac12\dot{\phi}^2+V(\phi) \, ,
\nonumber\\
P^{(\phi)} &\equiv&
\phantom{-}
{T^{i}_{i}}^{(\phi)} =
\frac12\dot{\phi}^2-V(\phi)\, ,
\label{defrhopphi}
\end{eqnarray}
and a part due to the non-minimal coupling, given from
Eq.~(\ref{Tphialpha}) as
\begin{eqnarray}
\rho^{(\alpha)} &\equiv& - {T^0_0}^{(\alpha)} =
- 6\alpha'H\dot{\phi} \,,
\nonumber\\
P^{(\alpha)} &\equiv&
\phantom{-}
{T^{i}_{i}}^{(\alpha)} =
2(\alpha'\ddot{\phi}+2H\alpha'\dot\phi+\alpha''\dot\phi^2)\,,
\label{defrhopalpha}
\end{eqnarray}
where $i = 1, \ldots ,3$ labels the spatial coordinates on the
brane.

The equation of motion (\ref{motiongeneralphi}) for the scalar
field, $\phi$, in the FRW geometry is
\begin{eqnarray}
\ddot{\phi}+3H\dot{\phi}+V'(\phi)=\alpha'R[g] \,,
\label{motionphi}\end{eqnarray}
where the intrinsic Ricci scalar for a FRW brane is
\begin{equation}
\label{Ricci}
R[g] = 6 \left( \dot{H} + 2H^2 +\frac{K}{a^2} \right) \,.
\end{equation}
In conventional 4D scalar-tensor gravity the Ricci scalar is often
eliminated from the scalar field equation of motion in favour of
the trace of the energy-momentum tensor, using the contracted
Einstein-Brans-Dicke equation for the Einstein tensor. In our
brane-world scenario the contracted effective Einstein equation
(\ref{Einsteineq}) yields a more complicated expression for the
Ricci scalar.

The non-minimal coupling of the scalar field to the Ricci curvature
on the brane through $\alpha(\phi)$ leads to the non-conservation of
the scalar field effective energy density
\begin{eqnarray}
\dot{\rho}+3H(\rho+P)=6\alpha'\dot\phi
\left(H^2+\frac{K}{a^2}\right). \label{conserrho}\end{eqnarray}
This equation can be deduced from the definition of $\rho$ and $P$
[see Eqs.~(\ref{redrho}), (\ref{redp}), (\ref{defrhopphi}),
(\ref{defrhopalpha})] and the equation of motion for $\phi$
(\ref{motionphi}). We see that $\rho$ and $P$ are conserved
whenever $\alpha$ is constant, i.e. when $\phi$  is a minimally
coupled scalar field. For this particular case, $\rho$ and $P$
reduce to $\rho^{(\phi)}$ and $P^{(\phi)}$ [see
Eq.~(\ref{defrhopphi})], respectively.

In general, although the
scalar field effective energy density $\rho$ is not
conserved, it is always possible to construct a total energy
density from the total energy momentum tensor $\tau_{\mu\nu}$, defined
in Eq.~(\ref{deftau}),
\begin{equation}
 \label{rhotot}
 \rho^{\rm tot} = \rho^{(\phi)} + \rho^{(\alpha)} - 6\alpha
\left( H^2 + \frac{K}{a^2} \right)
 \,,
 \end{equation}
which is locally conserved on the brane, in accordance with
Eq.~(\ref{conservationtau}).

\subsection{Low energy regime}

In order to analyse the different possible regimes for the
effective Friedmann equation on the brane, we introduce an
(arbitrary) constant brane tension $\sigma$, as in
Eq.~(\ref{bartau}) so that
\begin{equation}
\bar\rho = \rho - \sigma \,, \qquad \bar{P} = P + \sigma \,.
\end{equation}
If we then expand the quadratic term on the right-hand side of the
modified Friedmann equation (\ref{Friedmann}), we obtain
\begin{eqnarray}
 \label{quadexpansion}
 3 \left(H^2+\frac{K}{a^2}\right) =
  \Lambda_4 + \frac{\kappa_5^4\sigma}{6}
  \left[\bar\rho-6\alpha\left(H^2+\frac{K}{a^2}\right)\right]
 + \frac{\kappa^4_5}{12}\left[\bar\rho-6\alpha
 \left(H^2+\frac{K}{a^2}\right)\right]^2+\frac{C}{a^4}\;,
 \label{Friedmann2}
\end{eqnarray}
where $\Lambda_4$ is given by Eq.~(\ref{Lambda4}).

We can identify a low-energy regime corresponding to
\begin{equation}
\left| \bar\rho - 6\alpha\left(H^2+\frac{K}{a^2}\right) \right|
\ll \sigma
 \label{deflowrho}
 \,,
\end{equation}
where we recover from (\ref{quadexpansion}) an effective 4D
Friedmann equation
\begin{equation}
3 \left( 1+ \frac{\kappa_5^4\sigma\alpha}{3} \right)
 \left(H^2+\frac{K}{a^2}\right) \simeq
  \Lambda_4 + \frac{\kappa_5^4\sigma}{6} \bar\rho
 + \frac{C}{a^4}\,.
\label{low-Friedmann}
\end{equation}
with the effective gravitational coupling given by Eq.~(\ref{defPhilo}).

If we choose $\sigma=\sqrt{-6U}/\kappa_5$, i.e., set
$\Lambda_4=0$, and consider an anti-de Sitter bulk ($C=0$) then the constraint
equation (\ref{low-Friedmann}) allows us to express the low-energy
condition (\ref{deflowrho}) as
\begin{equation}
 \bar\rho \ll \sigma \left( 1 + \frac{\kappa_5^4\sigma\alpha}{3}
 \right) \,.
\end{equation}

\subsection{Strong coupling regime}

In order to identify the strong coupling regime we rewrite the
modified Friedmann (\ref{Friedmann}) equation as
\begin{equation}
 \label{Friedmannrewrite}
\left[ 1 - \frac{\rho}{6\alpha \left(H^2+\frac{K}{a^2}\right)}
\right]^2 = \frac{1}{\kappa_5^4\alpha^2(H^2+\frac{K}{a^2})} \left[
1 - \frac{\kappa_5^2U}{6(H^2+\frac{K}{a^2})} -
  \frac{C}{3a^4(H^2+\frac{K}{a^2})} \right] \,.
\end{equation}
We identify a strong coupling regime where
\begin{equation}
 \alpha^2 \gg \frac{1}{\kappa_5^4(H^2+\frac{K}{a^2})} \left[
 1 - \frac{\kappa_5^2U}{6(H^2+\frac{K}{a^2})} -
  \frac{C}{3a^4(H^2+\frac{K}{a^2})} \right] \,.
\end{equation}
in which case we recover from Eq.~(\ref{Friedmannrewrite}) an
effective 4D Friedmann equation
\begin{equation}
 \label{Friedmannhi}
6 \alpha \left(H^2+\frac{K}{a^2}\right) \simeq \rho \,,
\end{equation}
with the effective gravitational coupling given by Eq.~(\ref{defPhihi}).

Consistency of the last two equations implies that strong coupling
also requires a lower bound on the energy density
\begin{equation}
\rho\gg \frac{6}{\kappa_5^4\alpha}\left| 1- \frac{\kappa_4^2\alpha
U}{\rho}-\frac{2\alpha C}{a^4\rho}\right|.
\end{equation}

Note that the strong coupling form for the Friedmann equation,
(\ref{Friedmannhi}), can also be obtained from the low energy
regime, Eq.~(\ref{low-Friedmann}), for $\kappa_5^4\sigma\alpha \gg 1$
and $\Lambda_4=C=0$.

\subsection{Intermediate energy and weak coupling regime}

Having shown that one recovers two effectively 4D regimes in the
limits of low energy or strong coupling, it is interesting to
consider whether or not one can recover an essentially 5D regime
where $H^2\propto \rho^2$ as is found in Randall-Sundrum cosmology
(where $\alpha=0$) at high energies \cite{BDL,BDEL}.

The high-energy regime in the Randall-Sundrum model corresponds to
\begin{equation}
\rho \gg \sigma_{RS} \,,\label{conditionHEWC2}
\end{equation}
where $\sigma_{RS}=\sqrt{6|U|}/\kappa_5$ corresponds to the brane
tension required for a static Minkowski brane.
In the induced gravity model we must add the additional condition
\begin{equation}
\rho\gg \left |6\alpha
\left(H^2+\frac{K}{a^2}\right)\right| \,,\label{conditionHEWC1}
\end{equation}
Thus we require both high energy and weak coupling.
In this case, the modified Friedmann equation (\ref{Friedmann}) reads
\begin{equation}
 \label{Friedmann5D}
3\left(H^2+\frac{K}{a^2}\right) \simeq \frac{\kappa_5^4}{12}\rho^2
+\frac{C}{a^4}.
\end{equation}

Substituting Eq.~(\ref{Friedmann5D}) into the
inequality~(\ref{conditionHEWC1}) requires
\begin{equation}
\rho_-\ll\rho\ll\rho_+ \,,
\end{equation}
where
\begin{eqnarray}
\rho_{\pm}=\left| \frac{3}{\kappa_5^4\alpha} \left( 1 \pm
\sqrt{1-\frac{4\kappa_5^4\alpha^2C}{3a^4}}\right) \right|\,.
\end{eqnarray}
For this intermediate regime to exist requires both
\begin{equation}
\frac{\kappa_5^4\alpha^2C}{a^4} \ll 1 \,,
\end{equation}
and
\begin{equation}
 \label{LEWC}
\kappa_5^4|\alpha|\rho \ll 1 \,.
\end{equation}

Finally combining (\ref{conditionHEWC2}) and~(\ref{LEWC}) we
obtain the consistency condition
\begin{equation}
\sigma_{RS} \ll \rho \ll \frac{1}{\kappa_5^4|\alpha|} \,,
\end{equation}
which only exists for sufficiently weak coupling
\begin{equation}
 \label{suffweak}
|\alpha| \ll \frac{1}{\kappa_5^4\sigma_{RS}} \,.
\end{equation}

\section{De Sitter and Minkowski branes}

In this section, we describe some maximally symmetric branes that
can be obtained in the framework given in Sec.~\ref{III}. In
particular, we will obtain inflationary branes with de Sitter
geometry or purely Minkowski space-times on the brane. We consider
that the bulk is given by a 5D maximally symmetric space-time and
therefore the projected Weyl tensor on the brane is zero. For
simplicity we will use the spatially flat coordinate chart on the
brane so that $K=0$ and the Ricci scalar $R=12H^2=$constant.

{}From the Friedmann equation (\ref{Friedmann}) we see that we
require $\rho-6\alpha H^2=$constant. In addition, the last
condition and  the continuity equation (\ref{conserrho}) implies
that $P=-\rho$ for $H\neq0$. Note however that unlike 4D general
relativity, we do not necessarily require $\rho=$constant.

Equations~(\ref{redrho})-(\ref{redp}) for the density and
pressure of the non-minimally coupled scalar field give
\begin{equation}
 \label{rhoplusP}
\rho +P = (1+\alpha'')\dot\phi^2 + 2\alpha'\left( \ddot\phi -
 H\dot\phi \right) \,.
\end{equation}
The scalar field equation
(\ref{motionphi}) and the condition $P=-\rho$ then gives the
first-order constraint
\begin{eqnarray}
(1+2\alpha'')\dot\phi^2+ 2\alpha' \left(
12\alpha' H^2 - 4H\dot\phi - V'\right) =0 \,. \label{5pt1}
\end{eqnarray}
If the scalar field does not evolve in time ($\dot\phi=0$) and
$\alpha'\neq 0$, then we require $V'=12H^2\alpha'$, i.e. the potential
gradient is balanced by the non-minimal coupling to the scalar
curvature. The scalar field has to be at an extremum of the potential
($V'=0$) if $\phi$ is constant in time and $H=0$.

For a Minkowski brane $H=0$, the Raychaudhuri equation
(\ref{Raychaudhuri}) requires either $\rho+P=0$ or $\rho=0$. For
$\rho=0$ we must have $U=0$ from the Friedmann equation
(\ref{Friedmann}), but we may in principle have $P\neq0$.
Equation~(\ref{rhoplusP}) together with the equation of motion
(\ref{motionphi}) then yields for $\rho=0$
\begin{equation}
P = (1+2\alpha'')\dot\phi^2 - 2\alpha' V' \,.
\end{equation}
However, in the following we will restrict our discussion to de
Sitter or Minkowski branes with $P=-\rho$.

\subsection{De Sitter branes with $\dot\phi = 0$}

In the following, we will describe the fixed points of the theory,
i.e., values of the scalar field, $\phi=\phi_c$  such that
$\dot\phi=\ddot\phi=0$ and $\dot H=0$, where $H=H_c$ corresponds
to the Hubble parameter for $\phi=\phi_c$. For
$\dot\phi=\ddot\phi=0$ we necessarily have $\rho=-P=V_c$, where
$V_c=V(\phi)$.

Using Friedmann equation (\ref{Friedmann}) and the equation of motion of
the  scalar field (\ref{motionphi}), we obtain
\begin{eqnarray}
 \label{defHc}
H_c^2 &=& \frac{1}{6\alpha_c} \left\{ V_c +
  \frac{3}{\kappa_5^4\alpha_c} \left[ 1 \pm
    \sqrt{1-\frac23\kappa_5^4\alpha_c \left( \kappa_5^2 \alpha_c U -
        V_c \right)} \right] \right\},\label{6.1}\\
V'_c&=&12H_c^2\alpha'_c,\label{6.2}
\end{eqnarray}
where  $V'_c$, $\alpha_c$ and $\alpha'_c$ correspond to
 $V'(\phi_c)$, $\alpha(\phi_c)$ and $\alpha'(\phi_c)$,
respectively. Note that we require
$2\kappa_5^4\alpha_c(\kappa_5^2\alpha_cU-V_c)<3$ for $H^2$ to be
real.

To obtain a Minkowski brane with $H_c=0$ requires [see
Eq.~(\ref{Friedmann})] the usual Randall-Sundrum fine-tuning
between the 5D cosmological constant and the potential
\begin{equation}
V^2(\phi_c)=-\frac{6U}{\kappa_5^2} \geq 0 \,,
\label{Minkowskicond}
\end{equation}
In addition $\phi_c$ must coincide with an extremum of the
potential, $V_c'=0$. The Minkowski brane is only obtained for the
branch corresponding to the upper choice of sign in
Eq.~(\ref{defHc}) for $\kappa_5^4\alpha_c V_c+3 \leq0$ or lower
sign for $\kappa_5^4\alpha_c V_c+3 \geq 0$. Only for
$\alpha_c=-3/\kappa_5^4V_c$ do we obtain $H_c=0$ for both
branches.

In the following, we see under which conditions the fixed points
correspond to stable solutions. The potential  $V(\phi)$ and the
coupling $\alpha(\phi)$ can be approximated near $\phi_c$ by
\begin{eqnarray}
V(\phi)&\simeq&V_c+V'_c(\phi-\phi_c)+\frac12V''_c(\phi-\phi_c)^2,\\
\alpha(\phi)&\simeq&\alpha_c+\alpha'_c(\phi-\phi_c)+\frac12\alpha''_c(\phi-\phi_c)^2,
\end{eqnarray}
where $V''_c$ and $\alpha''_c$ are $V''(\phi_c)$ and $\alpha''(\phi_c)$,
respectively. The equation of motion (\ref{motionphi}) for
a small perturbation $\delta\phi=\phi-\phi_c$, to first-order
in $\delta\phi$ becomes
\begin{eqnarray}
 \label{deltaphieom0}
\delta\ddot\phi+3H_c\delta\dot\phi+(V''_c-12H_c^2\alpha''_c)\delta\phi=\alpha'_c\delta
R,
\end{eqnarray}
where $\delta R= R(\phi_c+\delta\phi)-12H_c^2$.
If $\delta R$ is negligible, we have that a
fixed point is stable when $V''_c-12H_c^2\alpha''_c>0$. However,
in general the perturbed Ricci scalar (\ref{Ricci}) can be calculated
using the Friedmann and Raychaudhuri equations (\ref{Friedmann}) and
(\ref{Raychaudhuri}), which gives
\begin{equation}
\delta R =
 -\frac{\kappa_5^4(V_c-6\alpha_cH_c^2)}
       {1+(1/3)\kappa_5^4\alpha_c(V_c-6\alpha_cH_c^2)}
\left[ \alpha'_c\delta\ddot\phi
+3\alpha'_cH_c\delta\dot\phi-4\alpha'_c H_c^2\delta\phi \right].
\end{equation}
Substituting this into the equation of motion of $\delta\phi$,
Eq.~(\ref{deltaphieom0}), we obtain (for
$\kappa_5^4(\alpha_c+3\alpha_c^{\prime2})(V_c-6\alpha_cH_c^2)\neq-3$)
the canonical equation of motion for a field perturbation in a de
Sitter background
\begin{eqnarray}
 \label{deltaphieom}
\delta\ddot\phi + 3H_c\delta\dot\phi + m_{\rm eff}^2 \delta\phi = 0 \,.
\end{eqnarray}
where the effective mass
\begin{equation}
m_{\rm eff}^2 \equiv \frac{V''_c-12H_c^2\alpha''_c-4\beta H_c^2}{{1+\beta}} > 0
 \,, \label{defDelta}
\end{equation}
with
\begin{equation}
\label{beta}
\beta = \frac{\kappa_5^4\alpha_c^{\prime2}(V_c-6\alpha_cH_c^2)}
             {1+(\alpha_c/3)\kappa_5^4(V_c-6\alpha_cH_c^2)} \,.
\end{equation}

The general solution to Eq.~(\ref{deltaphieom}) is given by
\begin{equation}
 \label{linear}
\delta\phi = C_+ \exp(\lambda_+t) + C_- \exp(\lambda_-t) \,.
\end{equation}
where
\begin{equation}
\lambda_\pm = \frac{-3H_c\pm\sqrt{9H_c^2-4m_{\rm eff}^2}}{2} \,.
\end{equation}
Thus the stability condition for the fixed point is simply $m_{\rm eff}^2>0$.
For a constant (minimal) coupling ($\alpha_c'=\alpha_c''=0$) this
stability condition takes the usual form $V''_c>0$. But non-minimal
coupling can stabilise the fixed point even for $V''_c<0$. For
example, for a Minkowski brane with $H_c=0$, the stability condition
$m_{\rm eff}^2>0$ reduces to $\phi_c$ being a maximum (or minimum) of the
potential for $1+\beta$ negative (or positive).

\paragraph{Quadratic model}

As an illustration, we apply the previous analysis to the following
simple quadratic model for the non-minimal coupling and potential:
\begin{eqnarray}
\alpha(\phi) &=&\alpha_0+\frac12\alpha_2\phi^2,\label{5.2} \\
V(\phi) &=&V_0+\frac12m^2\phi^2,\label{5.3}
\end{eqnarray}
where $\alpha_0$, $\alpha_2$, $V_0$ and $m^2$ are constants.

Imposing the condition (\ref{6.2}), yields two possible fixed points
for the model:
\begin{itemize}
\item $\phi_c=0$. The square of the Hubble parameter is given
by Eq.~(\ref{defHc}) with $V_c=V_0$ and $\alpha_c=\alpha_0$.
The parameter $\beta=0$ and the effective mass $m_{\rm eff}^2=
m^2-12\alpha_2H_c^2$. The fixed point is stable as long as
\begin{equation}
m^2 >
\label{stablem2}
 \frac{2\alpha_2}{\alpha_0} \left\{ V_0 +
  \frac{3}{\kappa_5^4\alpha_0} \left[ 1 \pm
    \sqrt{1-\frac23\kappa_5^4\alpha_0 \left( \kappa_5^2\alpha_0U - V_0
      \right)} \right] \right\} \,.
\end{equation}
\item $\phi_c\neq0$. This fixed point is obtained for $\alpha_2\neq0$
  when the Hubble
  constant satisfies $H_c^2=m^2/12\alpha_2$. The square of the Hubble
  parameter is given by Eq.~(\ref{defHc}) with $\alpha_0$ given by
$\alpha_0+(1/2)\alpha_2\phi_c^2$. {}From Eq.~(\ref{Friedmann}) we
obtain
\begin{eqnarray}
\phi_c^2 =
 \frac{2\alpha_0}{\alpha_2} - \frac{4V_0}{m^2}
 \pm \frac{4}{\kappa_5^2m^2}\sqrt{\frac{3m^2}{\alpha_2}+6\kappa_5^2U}
 \,. \nonumber
\end{eqnarray}
Thus this fixed point only exists for $m^2>-2\alpha_2\kappa_5^2U$.
{}From Eq.~(\ref{defDelta}) we find
$m_{\rm eff}^2=-4\beta H_c^2/(1+\beta)$ and hence this fixed
point is stable as long as $-1\leq \beta<0$.
\end{itemize}

\subsection{De Sitter branes with $\dot\phi\neq 0$}

We cannot find the general solution of the constraint Eq.~(\ref{5pt1})
for $\dot\phi\neq0$ without specifying the form of $V(\phi)$ and
$\alpha(\phi)$. Adopting the simple quadratic model
introduced in Eqs.~(\ref{5.2}) and  (\ref{5.3}) it can be shown
that there are solutions to the scalar field equation of motion
(\ref{motionphi}) with $\dot\phi\neq0$, which satisfy
Eq.~(\ref{5pt1}), when the mass of the scalar field satisfies
\begin{eqnarray}
\frac{m^2}{H^2}=
 \frac{2\alpha_2(1+6\alpha_2)(3+16\alpha_2)}{(1+4\alpha_2)^2} \,.
\label{5.6}
\end{eqnarray}
In this case we have a solution to the equation of motion
(\ref{motionphi}) where the scalar field evolves
exponentially with respect to cosmic time
\begin{equation}
\phi=\phi_0\exp\left(\mu Ht\right). \label{5.4}
\end{equation}
where the dimensionless parameter $\mu$ is given by
\begin{eqnarray}
\mu = \frac{2\alpha_2}{1+4\alpha_2}
 \,.\label{5.5}
\end{eqnarray}
For $-1/4<\alpha_2<0$ this describes the decay of the scalar field
to the fixed point with $\phi=0$, but for $\alpha_2>0$ or $<-1/4$
the $\phi=0$ fixed point is clearly unstable.

This limiting behaviour where $\phi\to0$ is consistent with linear
perturbations (\ref{linear}) studied in the previous sub-section
about the $\phi=0$ fixed point
for the particular case (\ref{5.6}).
The Hubble parameter, which remains constant for all $\phi$,
is thus given by the corresponding solution to Eq.~(\ref{defHc}) for
$\phi=0$:
\begin{equation}
H^2=\frac{1}{6\alpha_0}
        \left\{
V_0+\frac{3}{\kappa_5^4\alpha_0}
\left[1\pm\sqrt{1-\frac{2}{3}\kappa_5^4\alpha_0
\left(\kappa_5^2\alpha_0U - V_0\right)}\right]
        \right\},
\label{H0}
\end{equation}
which is real so long as
$2\kappa_5^4\alpha_0(\kappa_5^2\alpha_0U-V_0)<3$.

The effective energy density $\rho$ and the pressure $P$ of the
evolving scalar field, $\phi$, can be expressed as
\begin{equation}
\rho=-P=V_0+3\alpha_2 H^2 \phi^2\,,
\end{equation}
which will be time-dependent for $\alpha_2\neq0$.
On the other hand, it can be checked that,
\begin{equation}
\rho-6\alpha H^2 = V_0-6\alpha_0 H^2 = {\rm constant} \,.
 \label{5.8}\end{equation}

\section{Conformally coupled scalar field on the brane}

An interesting model to consider is the case of a conformally
coupled scalar field on the brane, with conformal coupling
\begin{equation}
\alpha=\alpha_0-\frac{1}{12}\phi^2,
\label{alphaconformal}\end{equation}
where $\alpha_0$ is a positive constant, and a vanishing
potential $V=0$.

It is known in 4D General Relativity that the trace of the
effective energy-momentum tensor of a conformally coupled scalar
field is zero. Therefore, the behaviour of a spatially homogeneous
conformally coupled field can be effectively described as a
radiation fluid \cite{HLBGGDMM}.  We will show that this remains
true in brane-world models.

We split the energy-momentum tensor (\ref{altdeftau}) as follows
\begin{equation}
 \label{conformaltau}
\tau_{\mu\nu} = {\widehat{T}}_{\mu\nu}
 -  2\alpha_0G_{\mu\nu} \,,
\end{equation}
where, from Eq.~(\ref{deftausplit}) we have
\begin{equation}
{\widehat{T}}_{\mu\nu}
 =
{T_{\mu\nu}}^{(\phi)}+
{T_{\mu\nu}}^{(\alpha)}+\frac{1}{6}\phi^2G_{\mu\nu}.
\label{defconformalT}
\end{equation}
The scalar field equation (\ref{motiongeneralphi}) then ensures that
$\widehat{T}_{\mu\nu}$ is traceless for the conformal coupling given
by Eq.~(\ref{alphaconformal}).
We recover the usual 4D result because we only use the scalar
field equation on the brane (\ref{motiongeneralphi}) and this
is formally the same as in 4D General Relativity case.
This result remains true if we include a quartic self-interaction
potential for the scalar $V=\lambda\phi^4$, but for simplicity we
will consider here the case of a non-self-interacting field
($\lambda=0$).

In the following, we will consider cosmological solutions where the
brane is homogeneous and isotropic.  For convenience, we define a
dimensionless scalar field $\chi=\phi/a$.
Now, the scalar field equation of motion (\ref{motionphi}) can be
rewritten as
\begin{equation}
\frac{d^2\chi}{d\eta^2}+K\chi=0,
\label{motionconformalphi}
\end{equation}
where $\eta=\int dt/a$ corresponds to the conformal time on the
brane and $K$ is the spatial curvature. In addition, we have that
the effective energy density $\widehat{\rho}$ and pressure
$\widehat{P}$, described by ${\widehat{T}}_{\mu\nu}$, are given by
\begin{eqnarray}
\widehat{\rho} = 3\widehat{P}
 =
 \frac{1}{2a^4}\left[\left(\frac{d\chi}{d\eta}\right)^2+K\chi^2\right]
 \,.
\label{rhopconformal}
\end{eqnarray}
Using Eq.~(\ref{rhopconformal}) and the first integral of
the scalar field equation of motion (\ref{motionconformalphi}),
we obtain $\widehat{\rho}=B/a^4$, where $B$ is an
integration constant.

If in addition, we consider a non-vanishing, but constant potential
$V=V_0$ on the brane, we have that ${\widehat{T}}_{\mu\nu}$ is no
longer trace free. On the other hand, the
evolution of the scalar field $\chi$ is unchanged and given by
Eq.~(\ref{motionconformalphi}), while $\widehat{\rho}$ and
$\widehat{P}$ are shifted such that
\begin{eqnarray}
\widehat{\rho}&=& \frac{B}{\phantom{3}a^4}+V_0,\\
\widehat{P}&=& \frac{B}{3a^4}-V_0.
\end{eqnarray}

The cosmological evolution of the brane is given by
Eqs.~(\ref{Friedmann}) and (\ref{Raychaudhuri}), where now $\rho$,
$P$ and $\alpha$ are substituted by $\widehat{\rho}$,
$\widehat{P}$ and $\alpha_0$, respectively. The Hubble parameter
(\ref{Hubble}) thus reads
\begin{eqnarray}
H^2 = -\frac{K}{a^2} + \frac{1}{6\alpha_0} \left\{
  \frac{B}{\phantom{3}a^4} + V_0+\frac{3}{\kappa^4_5\alpha_0} \left[1
\pm\sqrt{1 + \frac{2}{3}\kappa_5^4\alpha_0 \left( - \kappa_5^2
    \alpha_0 U + V_0 +\frac{B-2\alpha_0C)}{a^4}\right)}\,
\right]\right\}\;, \label{conformalFriedmann}\end{eqnarray}
This Friedmann equation is the same as that found in \cite{MMT}
for a radiation filled brane-world universe with a non vanishing
brane tension.

We note that it is possible to recover the conventional evolution for
a 4D cosmology filled with radiation and an effective vacuum energy
density with a fine tuning of the parameters of the solution. This is
possible, when the energy density of the conformally coupled scalar
field on the brane exactly cancels the effect of the projected Weyl
tensor on the brane, i.e.,
\begin{equation}
B=2\alpha_0 C \,.
\end{equation}
Then the Friedmann equation (\ref{conformalFriedmann}) becomes
\begin{equation}
3\left(H^2+\frac{K}{a^2}\right)=\frac{B}{2\alpha_0
a^4}+\Lambda_{\pm},
\end{equation}
where the effective cosmological constant is given by
\begin{equation}
\Lambda_{\pm}=
\frac{1}{2\alpha_0} \left\{V_0 + \frac{3}{\kappa_5^4}
  \left[1\pm\sqrt{1-\frac23\kappa_5^4\alpha_0(\kappa_5^2\alpha_0 U
      -V_0)} \right] \right\} \,.
\end{equation}
A vanishing cosmological constant on the brane requires the usual
Randall-Sundrum fine-tuning $V_0^2=-6U/\kappa_5^2$. We then obtain
$\Lambda_\pm=0$ for $\pm(\kappa_5^4\alpha_0 V_0)\leq0$.

Going back to the projected Einstein equations (\ref{Einsteineq}) we
can see that it is possible to recover the standard 4D Einstein
equations
\begin{equation}
G_{\mu\nu} = -\Lambda_4 g_{\mu\nu} + \frac{1}{2\alpha_0}
T_{\mu\nu}^{\rm c.c.} \,,
\end{equation}
for the special case of a conformally coupled field with (trace-free)
energy-momentum tensor if it exactly matches the projected Weyl
tensor
\begin{equation}
T_{\mu\nu}^{\rm c.c.} = -2\alpha_0 E_{\mu\nu} \,.
\end{equation}
In this case the conformally coupled field and the projected Weyl
tensor exactly cancel out in the total effective energy-momentum
tensor (\ref{conformaltau}) on the brane,
$\tau_{\mu\nu}=(2\alpha_0\Lambda_4-V_0)g_{\mu\nu}$, and hence
$\Pi_{\mu\nu}\propto g_{\mu\nu}$ in Eq.~(\ref{quadratic}). The
conformally coupled energy-momentum tensor, $T_{\mu\nu}^{\rm c.c.}$
(or equivalently the projected Weyl tensor, $E_{\mu\nu}$) then only
appears {\em linearly} in the induced Einstein equations
(\ref{Einsteineq}).

\section{Discussion}

In this paper we have studied the field equations for a scalar
field living on a 4D brane embedded in 5D vacuum space-time,
including the effect of a non-minimal coupling of the field to the
4D scalar curvature on the brane. This is a natural generalisation
of previous studies of the dynamics of minimally coupled scalar
fields on the brane, just as Brans-Dicke scalar-tensor models are
a natural generalisation of minimally coupled fields in 4D general
relativity. Such a non-minimal coupling would be expected to arise
as a quantum correction for any self-gravitating field, but in the
present paper we have just considered the classical dynamics of an
effective theory with non-minimal coupling.

In a 4D scalar-tensor gravity theory with a non-minimally coupled
scalar field it is always possible to perform a conformal
transformation \cite{wald} $g_{\mu\nu}\to \Omega^2(\phi) g_{\mu\nu}$
to recast the theory as Einstein gravity plus a minimally coupled
field, in what is known as the Einstein frame \cite{Maeda}. This is no
longer possible in a brane world context with a non-minimally coupled
scalar field on the brane, as the bulk gravity already defines a ``5D
Einstein frame'' \cite{MW}. The non-minimal coupling of the scalar on
the brane results in the Einstein-Brans-Dicke tensor
(\ref{EinsteinBD}) appearing as a source term in the total
energy-momentum tensor on the brane. If one attempts to simplify this
by a conformal transformation to the ``4D Einstein frame'' on the
brane, then this simplifies the total energy-momentum source term
on the brane, but results in more complicated effective gravitational
field equations in the bulk. There seems to be no easy way to avoid
the rather messy gravitational field equations for a non-minimally
coupled scalar field on the brane.

We identify two different regimes in which the evolution reduces to
the usual 4D form. At low energies (relative to the brane tension
$\sigma$) the projected 5D Einstein equations reduce to an effective
4D gravity theory (\ref{Einsteinlo}), which is a generalisation of the
Randall-Sundrum model \cite{RS2}. The non-minimal coupling
$\alpha(\phi)$ leads to a correction to the effective gravitational
constant on the brane (\ref{defPhilo}). On the other hand, if the
non-minimal coupling term is large so that the effects of the bulk
gravity is negligible, we recover an effective 4D scalar-tensor
gravity theory (\ref{Einsteinhi}) where $\alpha(\phi)$ describes the
gravitational coupling (\ref{defPhihi}).

We give the form of the modified Friedmann equation for
homogeneous and isotropic cosmologies with a non-minimally coupled
scalar field. For a FRW brane moving in 5D anti-de Sitter
space-time it is then possible to give expressions for the 4D
low-energy and strong-coupling regimes in terms of the energy
density. Only for sufficiently weak coupling (\ref{suffweak}) is
it possible to recover an intermediate ``5D'' regime where the
Hubble expansion is linearly proportional to the scalar field
energy density on the brane \cite{BDL}.

We have given the projected field equations on the brane following
the approach of Shiromizu, Maeda and Sasaki \cite{SMS} where the
non-local effect of bulk gravity is described by the projection of
the 5D Weyl tensor. The most general 5D vacuum solution respecting
the symmetries of a homogeneous and isotropic (FRW) brane is 5D
Schwarzschild anti-de Sitter where the projected Weyl tensor acts
like a radiation fluid.

An interesting special case is that of a conformally coupled scalar
field on the brane. As in 4D gravity, one can use the scalar field
equation of motion to define a trace-free energy-momentum tensor
(\ref{defconformalT}) for a conformally coupled field on the brane. In
general this obeys the same modified Friedmann equation as found
previously \cite{Deffayet,MMT} for a radiation fluid on a brane with
fixed induced gravity coupling $\alpha_0$. But for particular values
of the conformal field's energy density it is possible for it to
exactly cancel out the non-local effect from the projected Weyl tensor
and we recover a standard 4D Friedmann equation for a conformal field.

We also identify de Sitter brane solutions with constant $H$. We find
solutions with a constant scalar field displaced from the minimum of
the potential, where the potential gradient is balanced by the
gradient of the non-minimal coupling term. But for some scalar field
Lagrangians it is also possible to find de Sitter solutions with
constant 4D Ricci scalar, but non-constant scalar field.

It is natural to consider extending previous analyses of slow-roll
inflation due to a self-interacting scalar field on the brane
\cite{chaotic} to include the effect of a non-minimal coupling for the
scalar field to the induced Ricci curvature on the brane. Several
authors have considered the spectrum of scalar metric perturbations
produced by quantum fluctuations of an inflaton field on the brane in
the presence of a constant induced gravity correction
\cite{lefteris,BLMW}. Indeed we have recently shown that the 4D
consistency relation for the tensor-scalar ratio from inflation
remains true with a constant induced gravity correction. It would be
interesting to see whether this remains true for a scalar field with
non-minimal coupling $\alpha(\phi)$. However our ability to relate the
scalar metric perturbations produced during inflation to
observables at late times may be limited due to the non-conservation
of the scalar field energy density $\rho$ in Eq.~(\ref{defrhopphi}). Only
the total effective energy density $\rho^{\rm tot}$ in
Eq.~(\ref{rhotot}) is locally conserved and so we require strictly
adiabatic perturbations in this total effective energy density in
order for the scalar curvature perturbation to remain constant in the
large scale limit \cite{WMLL}. We leave this interesting question for
future work.

\acknowledgments

We thank Carlos Barcelo and Kei-ichi Maeda for helpful discussions.
MBL is funded by the Spanish Ministry of Education, Culture and Sport
(MECD), and also partly supported by DGICYT under Research Project
BMF2002 03758. DW is supported by the Royal Society.

\end{document}